\documentstyle[12pt,epsf]{article}
\begin{document}
\begin{center}
{\ {\Large {\bf Azimuthally Sensitive HBT Analysis}}}\\

\vskip 0.6cm

{\sl Peter Filip\footnote{Permanent address: Institute of Physics, 
SAS, Bratislava, SK-842 28}}\\
\vskip 0.25cm
Based on talk given at Lawrence Berkeley Laboratory \\
1 Cyclotron Road, Berkeley, CA 94720, USA \\
(16 July 1996) \\

\vspace{0.62cm}
\end{center}

\vspace{0.5cm}
\begin{abstract}

\vspace{0.3cm} 
In this paper we investigate a possibility to perform shape-sensitive
HBT analysis in transversal plane. We study presently used 
Side-Out decomposition of the correlation function and we do not find it 
to be very sensitive for the detection of a shape of source distribution in 
transversal plane.
Therefore we suggest to use X-Y decomposition of the correlation
function for the HBT analysis of non-central heavy ion collisions.
We test the method suggested on artificial and non-central 
Pb-Pb 158GeV/n events produced
by CASCUS event generator.
Further possible enhancement of the method is sketched.
\newline 

\end{abstract}

\vspace{0.2cm}

\begin{center}
{\bf 1. Introduction}
\end{center}

HBT interferometry analysis has become a powerful method in present
day heavy ion collision (HIC) experiments. Nice theoretical and experimental
work has been done in the field of HBT analysis of central HIC events. 
Long-Side-Out decomposition of correlation function enabled us to separate
space and time properties of source distribution of bosons $\rho (\vec x,t)$.
Even more sophisticated models of emission functions containing flow and
velocity parameters have been analyzed in various coordinate systems 
\cite{Heinz1}. 

On the other hand HBT analysis is still apparently not fully accomplished
technique. For example Gamov factor correction  
which was expected to describe
influence of coulomb final state interactions \cite{Bowler}
on the correlation function
was found to be in disagreement with recent experimental results \cite{QM96}.

Recently HBT analysis of non-central HIC events has been performed on data
taken at SPS \cite{QM96}
and AGS \cite{QM95}. Extracted radius 
parameters were found to be dependent on centrality of collision and also
on azimuthal orientation of the momentum of pions in respect to the
reaction plane \cite{QM95}. For this analysis of non-central events a methods
originally developed for cylindrically symmetric systems were used.

We think that a more comprehensive information about 
the source distribution $\rho (\vec x,t)$ 
can be obtained. The purpose of this paper is to 
introduce an HBT analysis method which is suitable for the study of 
geometrical properties of source distributions e.g. a shape or structure
of $\rho (\vec x)$.

This paper is organized as follows: In Section 2 we study Side-Out 
decomposition of correlation function and we find it to be unsensitive
to the shape of the source distribution in transversal plane. In Section 3
we introduce X-Y decomposition of correlation function and we show it to be
suitable for the study of the geometrical properties of source 
distribution. A possibility to perform azimuthally sensitive HBT analysis
of non-central events in the case when orientation of impact parameter 
$\vec b$
is not known is investigated in Section 4. In subsequent sections we study
influence of $\vec x - \vec p$ correlation of source distribution 
$\rho (\vec x,t)$ on
transversal HBT analysis and also possible further enhancement of the 
HBT interferometry method.

\vskip0.5cm
\begin{center}
{\bf 2. Side-Out Decomposition of Correlation Function}
\end{center}

In this section we shall recall advantages of the 
Side-Out decomposition of correlation function. We start with
simplified expression for the correlation function produced by
$\vec x - \vec p$ non-correlated source $\rho (\vec x,t)$. Assuming
that probability of emission of a single boson does not depend
on momentum and that emission of bosons is incoherent we write:
$$
C(\Delta \vec p)=\int \int d^3\vec x_1 d^3\vec x_2 \Big [
                 \rho(\vec x_1,t_1). \rho (\vec x_2,t_2) 
		 |\Psi _{12}|^2 \Big ]
                 dt_1 dt_2
\eqno{(1)}
$$

where

$$
|\Psi _{12}|^2 = 
{\textstyle \frac{1}{2}} 
\Big|e^{ip_1 x_1}e^{ip_2 x_2} + e^{ip_2 x_1}e^{ip_1 x_2} \Big|^2
= 1 + \cos (\Delta \vec x \cdot \Delta \vec p - \Delta t \cdot \Delta E)
\eqno{(2)}
$$
By simple kinematic relations we can prove \cite{Zajc}:
$$
\Delta E = \Delta \vec p \cdot \vec \beta _{out} \quad ; \quad
\vec \beta _{out} = \frac{\vec p_1 + \vec p_2}{E_1 + E_2} \quad , \quad
\Delta \vec p = \vec p_1 - \vec p_2
\eqno{(3)}
$$
Thus decomposing scalar product in Eq.(2) into Out and Side orthogonal
components we can separate time information about the emission source from the
Side component of correlation function:
$$
1+\cos (\Delta \vec p_{side} \cdot \Delta \vec x_{side} +
\Delta \vec p_{out} \cdot [ \Delta \vec x_{out} - 
\Delta t \cdot \vec \beta _{out}]
)
\eqno{(4)}
$$

This Side-Out decomposition of correlation function allows to study time
properties of source $\rho (\vec x,t)$ as a differnce of $R_{out},R_{side}$
extracted parameters. One of the main messages contained in this paper is:

\vskip0.3cm
\begin{center}
\begin{tabular}{|c|}
\hline
{\it Side-Out decomposition of correlation function  destroys information}\\
{\it about possible transversal shape of source distribution 
$\rho (\vec x,t)$.}
\\
\hline
\end{tabular}
\end{center}
\vskip0.01cm 

Also for the sake of simplicity we concentrate on the study of transversal 
properties of source distribution $\rho (\vec x,t)$ in this paper. 
For this purpose we shall 
investigate  properties of {\it transversal} correlation function
constructed from pairs with very small longitudinal component of relative
momentum  $|\Delta p_l | \ll |\Delta \vec p_t |$. Then we can
neglect longitudinal term in scalar product 
$\Delta \vec p \cdot \Delta \vec x $ in equation (2):
$$
\Delta \vec p \cdot \Delta  \vec x =
\Delta p_l \cdot \Delta x_l + \Delta \vec p_t \cdot \Delta \vec x_t
\approx
\Delta \vec p_t \cdot \Delta \vec x_t
\eqno{(5)}
$$
and for the transversal correlation function we write:
$$
C^T(\Delta \vec p_t)= \int \int d^2\vec x_1 d^2\vec x_2 \Big[ 
\rho ^t (\vec x_1,t) . \rho ^t (\vec x_2,t) 
\big[ 1+\cos(\Delta \vec p_t \cdot \Delta \vec x_t - \Delta t \cdot \Delta E) 
\big] \Big]
dt_1 dt_2
\eqno{(6)} 
$$
Source distribution $\rho ^t (\vec x)$ in (4) is a projection
of three-dimensional source distribution $\rho (\vec x)$ onto the transversal
plane: $\rho ^t (\vec x_t) = \int \rho (\vec x_t, x_l) dx_l $.

On Fig.1 we show transversal correlation function in Side-Out decomposition
for artificially asymmetric S-Pb events. Momenta of bosons were generated in
agreement with experimentally measured $p_t$ and rapidity distributions of
pions in central S-Pb 200GeV/n collisions in the same way as it was done in
computer simulation \cite{Humanic}. Space distribution of emission points of
pions in transversal plane was changed artificially as it is shown on Fig.1a.

\vskip0.5cm
\centerline{\epsfxsize=11.7cm\epsffile{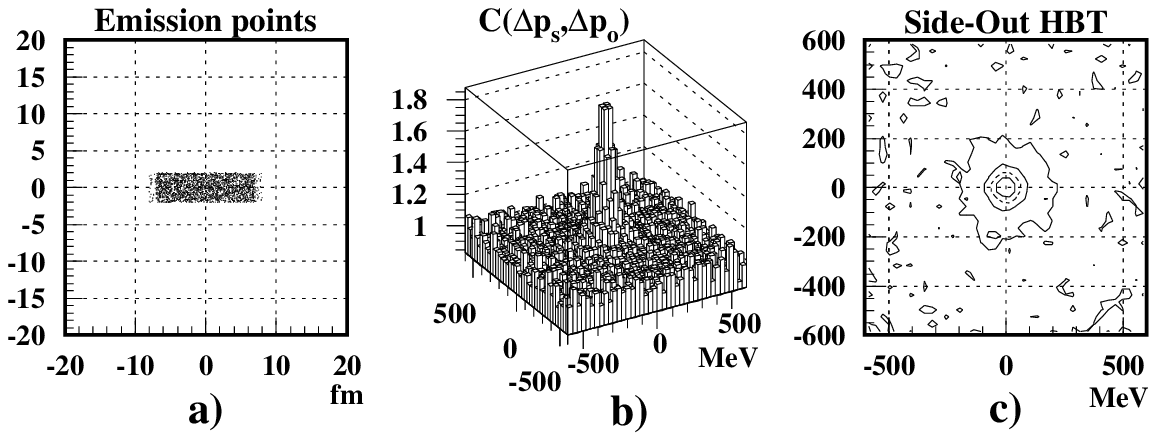}}
\vskip0.2cm
\centerline{\parbox{12cm} {\small {\bf Fig.1} Side - Out Correlation
function for asymmetric source in transversal plane. Pions are assumed
to be emitted at the same time ($\Delta t = 0)$.}}
\vskip0.2cm

Transversal correlation function $C^T(\Delta p^t_{side},\Delta p^t_{out})$
does not exhibit any azimuthal asymmetry as a consequence of asymmetrical 
shape of source distribution in Fig.1a - it is azimuthally non-sensitive.
In the next section we show that another decomposition of correlation
function is sensitive to the shape properties of $\rho (\vec x,t)$.
Here we shall try to explain what is a mechanizm of destroying the
shape information in Side-Out decomposition analysis.

\vskip0.2cm
\centerline{\epsfxsize=9.7cm\epsffile{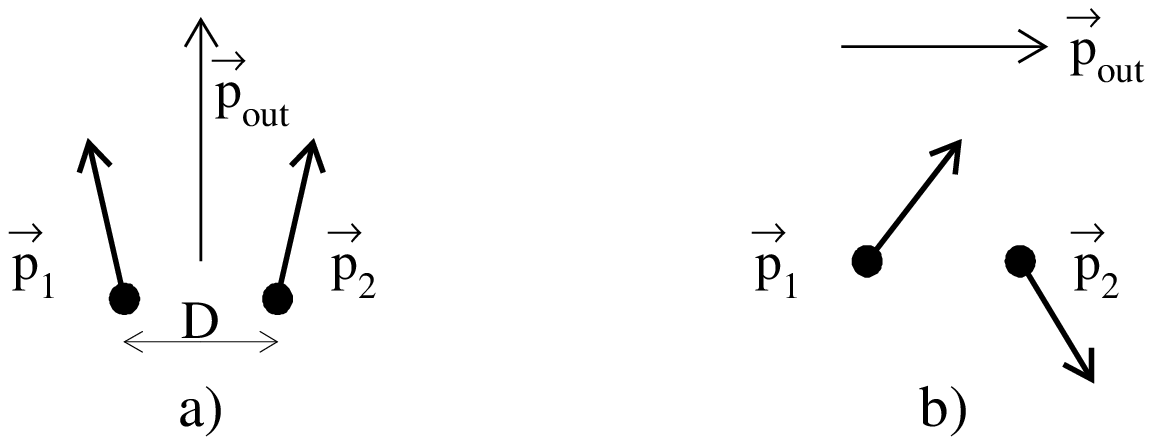}}
\vskip0.2cm
\centerline{\parbox{12cm} {\small {\bf Fig.2} Two-point source
emitting bosons. Relative momentum $\Delta \vec p$ 
may be of different orientation than it is shown here. }}
\vskip0.4cm

Let us study a very simple case -- static two-point source emitting
pairs of bosons simultaneously (Fig.2). If bosons are emitted as it is shown 
on Fig.2a then separation of emission points in Out and Side directions is:
$$
\Delta x_{out} = 0 \quad ; \quad \Delta x_{side} = D
\eqno{(7)}
$$
On Fig.2b the orientation of sum $\vec p_1 + \vec p_2$ leads to
a totally different situation:
$$
\Delta x_{out} = D \quad ; \quad \Delta x_{side} = 0
\eqno{(8)}
$$
Since the orientation of vector 
$\vec p_{out}=\frac{\vec p_1+\vec p_2}{E_1+E_2}$ is random values 
of $\Delta x_{out}, \Delta x_{side}$ are randomly distributed but
fulfilling condition $\Delta x_{out}^2 + \Delta x_{side}^2 =D^2$. It means
that source behaves as a {\it rotating} in Side-Out transversal
HBT analysis.

In the next section we introduce X-Y approach to HBT which does not
exhibit this feature and 
we find it to be sensitive to a non-trivial geometrical structure
of source distribution in transversal plane.

\vskip0.5cm
\begin{center}
{\bf 3. X-Y Decomposition of Transversal Correlation Function}
\end{center}

Let us imagine that orientation of impact parameter $\vec b$ is known
with rather high precision for each non-central HIC event. In this case
we can rotate measured  momenta of particles in all events to have the same 
orientation of impact parameter.

Because of asymmetrical overlapping region of colliding ions in non-central
events the  shape of 
source distribution is asymmetrical in transversal plane. This shape
information would be destroyed if Side-Out type of HBT analysis without
any additional cuts was applied\footnote{See Section 5 for a possibility
to apply directional cuts in momentum space.}.

On Fig.3 we show transversal correlation function in X-Y decomposition for
the same source distribution as shown on Fig.1a. Correlation function 
$C^T(\Delta p_x, \Delta p_y )$ is clearly asymmetrical and this reflects
the shape of original source distribution $\rho (\vec x,t)$. 

\vskip0.2cm
\centerline{\epsfxsize=11.7cm\epsffile{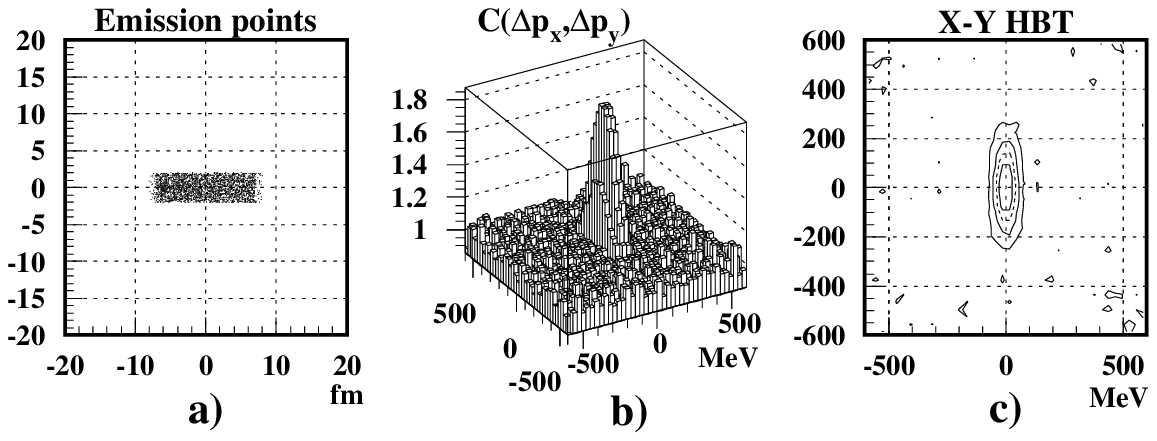}}
\vskip0.1cm
\centerline{\parbox{12cm} {\small {\bf Fig.3} X-Y transversal 
HBT analysis. Fig.3c) is a contourplot
of Fig.3b). See also Fig.1 for comparison.
}}
\vskip0.2cm

In the first 
approximation we suggest to use following gaussian parametrization
of the correlation function:
$$
C^T(\Delta p_x,\Delta p_y)=1+\lambda . 
e^{-\Delta p_x^2.R_x^2/2-\Delta p_y^2.R_y^2/2}
\eqno{(9)}
$$
which corresponds to the static (sharp freeze-out) source distribution
$$
\rho ^T (x,y)=c . e^{-x^2/R_x^2 - y^2/R_y^2}
\eqno(10)
$$

Results of this simple X-Y decomposition HBT analysis for Pb+Pb non-central 
158 GeV/n HIC events generated by CASCUS generator are presented
in next section. Here we shall discuss the question of lifetime of source
which was the main reason for Side-Out decomposition of $C(\Delta \vec p)$
as it is described in the preceeding section.

Scalar product in the argument of $\cos$ in equation (2) in X-Y coordinate
system ($\Delta p_z \approx 0$) is:
$$
\cos (\Delta x \cdot \Delta p_x + \Delta y \cdot \Delta p_y - 
\Delta t \cdot \Delta E )
\eqno{(11)}
$$

It means that time information contained in $\rho (\vec x,t)$ is mixed in the
same way with $R_x$ and $R_y$ parameters of the source extracted by fit of the
experimentally measured correlation function to equation (9). 
Consequently  if source 
$\rho (\vec x,t)$ is asymmetrical in X-Y coordinate system  then size 
parameters $R_x,R_y$  extracted by the fit should still exhibit asymmetry
$R_x \neq R_y$. 

Moreover if we rely on the argument of
{\it cos} in equation (2) then influence time dependet factor 
$\Delta t \cdot \Delta E$ on our $C^T(\Delta p_x,\Delta p_y)$ 
formally vanishes if only
pairs fulfilling the following condition
$$
\Delta E = E_1 - E_2 \approx 0 \quad 
\Longrightarrow \quad 
|\vec p_1|\approx |\vec p_2|
\eqno{(12)}
$$
are selected for HBT analysis. Condition (12) together with our previous
requirement $\Delta p_z \approx 0$ still leaves enough freedom in the 
distribution of pairs in relative momentum space $(\Delta p_x, \Delta p_y)$. 

We finish this section by schematic diagram depicting advantages of the
above described Side-Out and X-Y HBT methods.

\vskip0.3cm
\centerline{\epsfxsize=9.7cm\epsffile{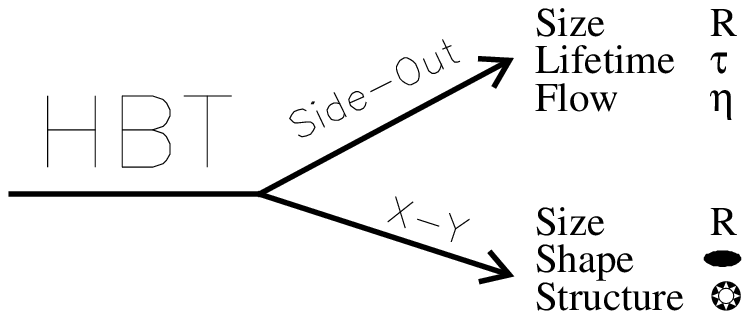}}
\vskip0.0cm
\centerline{\parbox{10cm} {\small {\bf Fig.4} Advantages of Side-Out
and X-Y HBT analysis.}}
\vskip0.1cm

Side-Out decomposition of $C(\Delta \vec p)$ allows to study mainly lifetime
$\tau$ and flow $\eta$ parameters of source (besides total size $R$). 
X-Y decomposition is sensitive to the shape and possibly other
non-trivial geometrical structure of source distribution $\rho (\vec x)$.

\vskip0.5cm
\begin{center}
{\bf 4. Pb+Pb 158 GeV/n Non-central Events CASCUS Simulation}
\end{center}

Azimuthally sensitive X-Y approach described in preceeding
section was tested using a computer simulation of Pb+Pb 158GeV/n non-central
events $|\vec b|=7fm$. Main structure of the simulation is shown
on Fig.5.

\vskip0.4cm
\centerline{\epsfxsize=11cm\epsffile{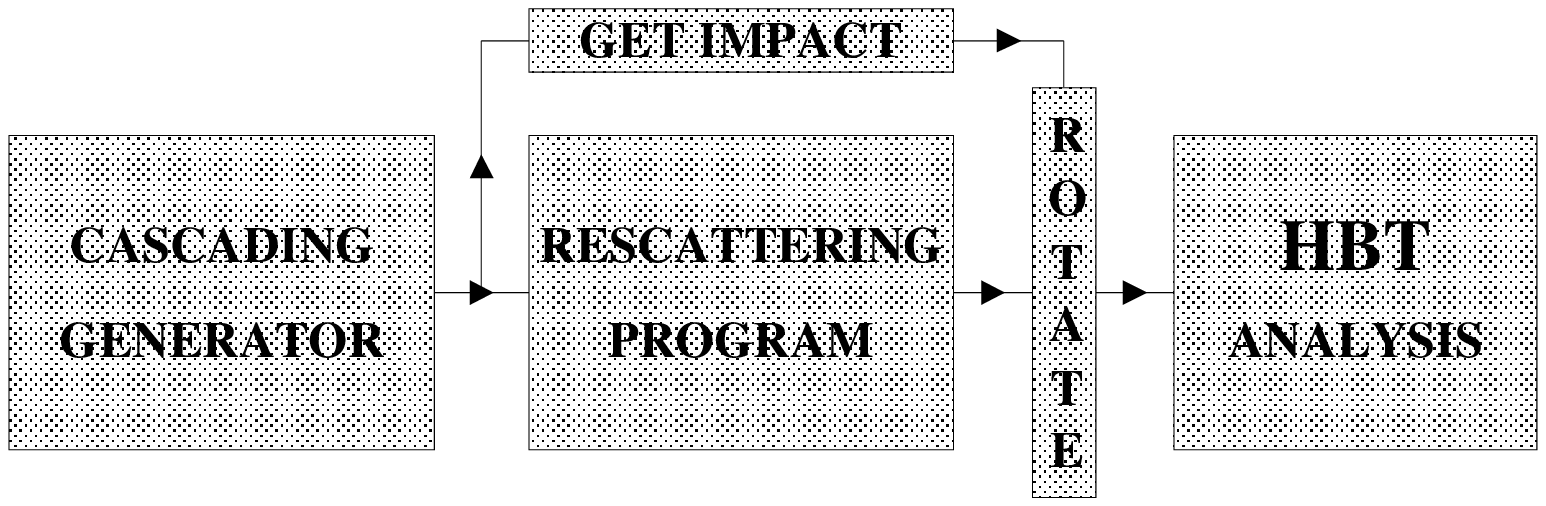}}
\vskip0.3cm
\centerline{\parbox{12cm} {\small {\bf Fig.5} CASCUS generator setup
used for the simulation of non-central Pb+Pb 158 GeV/n HIC events. }}
\vskip0.4cm

Initial positions and momenta of pions created in non-central Pb+Pb
collisions were produced by Cascading Generator \cite{Zavada}. 
Pions were evolved in time using enhanced version \cite{Acta} of
rescattering program \cite{Humanic}. Resulting distribution
of last interactions of pions was used for HBT analysis.

Events with random
orientation of impact parameter $|\vec b| =7fm$ were rotated to have
the same orientation of impact parameter $\vec b$.
Cascading generator program \cite{Zavada} was run at Czech Academy of
Sciences (CAS) and rescattering program \cite{Acta} was written at 
Comenius University in Slovakia (CUS). 

Correlation function was constructed in the way
similar to that used in simulation \cite{Humanic} but in X-Y coordinate
system. Result was fitted to analytical expression (9). 
Parameters $R_x,R_y$ obtained by the fit are 
summarized in the following table.

\vskip0.4cm
\begin{center}
\begin{tabular}{|c|c|c|c|}
\hline
X-Y HBT & $\rho (\vec x)$ & BEFORE  & AFTER  \\
RESULTS & ORIGINAL & RESCATTERING & RESCATTERING \\
\hline 
TOY    & $S_x=16fm$ & $R_x=8.76\pm 0.02fm$ & $R_x=8.12\pm 0.02fm$ \\
\mbox{\epsfxsize=0.9cm\epsffile{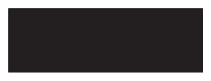}}
     & $S_y=4fm$  & $R_y=2.35\pm 0.02fm$ & $R_y=3.67\pm 0.02fm$ \\
EVENTS & $S_x-S_y=12fm$ & $\Delta R=8.4fm$ & $\Delta R=7.2fm$\\
\hline
Pb + Pb& $S_x=11fm$ & $R_x=5.28\pm 0.05fm$ & $R_x=5.48\pm 0.05fm$ \\
\mbox{\epsfxsize=0.9cm\epsffile{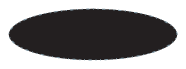}}
  & $S_y=6fm$ & $R_y=3.44\pm 0.05fm$ & $R_y=4.64\pm 0.05fm$ \\
EVENTS & $S_x-S_y=9fm$ & $\Delta R=4.0fm$ & $\Delta R=2.9 fm$\\
\hline
\end{tabular}
\end{center}
\centerline{\parbox{12cm} {\small {\bf Tab.1} Results of X-Y HBT 
analysis. $R_x,R_y$ parameters are extracted by fit to Eq.(9).
$S_x,S_y$ are size parameters of source distribution
$\rho (\vec x)$. }}
\vskip0.3cm

As a consequence of asymmetrical shape of source distribution $\rho (\vec x)$
extracted
parameters $R_x,R_y$ are not of the same value what is demonstrated
also by non-zero value of $\Delta R = \sqrt {R^2_x - R^2_y}$ 
in the above table.

Comparing  results obtained by  HBT analysis of the events before and
after rescattering proces we conclude that rescattering process
smears the shape of the original source distribution.
On the other hand $\vec x - \vec p $ correlation in the resulting 
source distribution $\rho (\vec x, \vec p)$ after the rescattering
process allows to perform another type of shape sensitive analysis 
based on directional cuts in momentum space (see next section).

\begin{center}
{\bf 5. Influence of $\vec x - \vec p$\, Correlation}
\end{center}

Let us imagine that there is no $\vec x - \vec p$ correlation in 
the distribution of pions before the rescattering process.
It means we assume that pions are created in individual nucleon - nucleon
collisions and that there is no collective behaviour of nucloens before
the emission of pions. Then geometrical properties of the emission points 
distribution
of pions $\bar \rho (\vec x,\vec p) = \rho (\vec x). \rho (\vec p)$ do not 
depend on cuts in momentum space. 
On the other hand rescattering process creates $\vec x - \vec p$ 
correlation in last interactions distribution of pions 
$\rho ^{LI}(\vec x, \vec p)$ (see Fig.5).

\vskip0.2cm
\centerline{\epsfxsize=11.7cm\epsffile{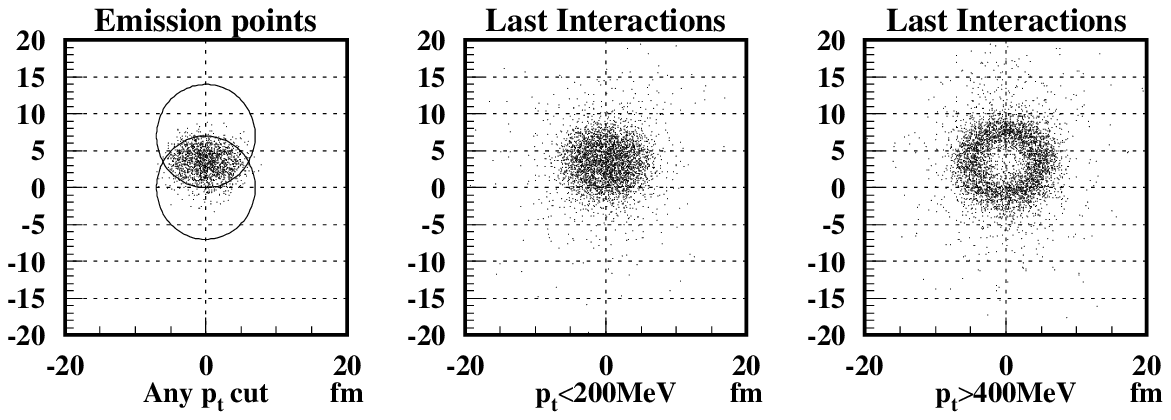}}
\vskip0.2cm
\centerline{\parbox{12cm} {\small {\bf Fig.5} Emission points of
pions (CASCUS Pb+Pb 160 GeV/n ; $\vec b = 7fm$) in transversal plane. 
Distribution
of Last Interactions depends on $p_t$ cut.}}
\vskip0.2cm

Distribution of last interactions depends on absolute value of
transversal momentum. This explains dependence of extracted
HBT radii parameters on $p_t$ which was measured experimentally \cite{NA44}.
Moreover rescattering process creates also 
a directional $\vec x - \vec n_p$ type of correlation \cite{UAW}.
This type of correlation can also be used for detection of the transversal 
shape of $\rho ^{LI} (\vec x, \vec p )$ distribution. On Fig.6 we show
$\rho ^{LI} (\vec x, \vec p )$ distribution of pions for selected directional
cuts in transversal momentum space.

\vskip0.2cm
\centerline{\epsfxsize=11.4cm\epsffile{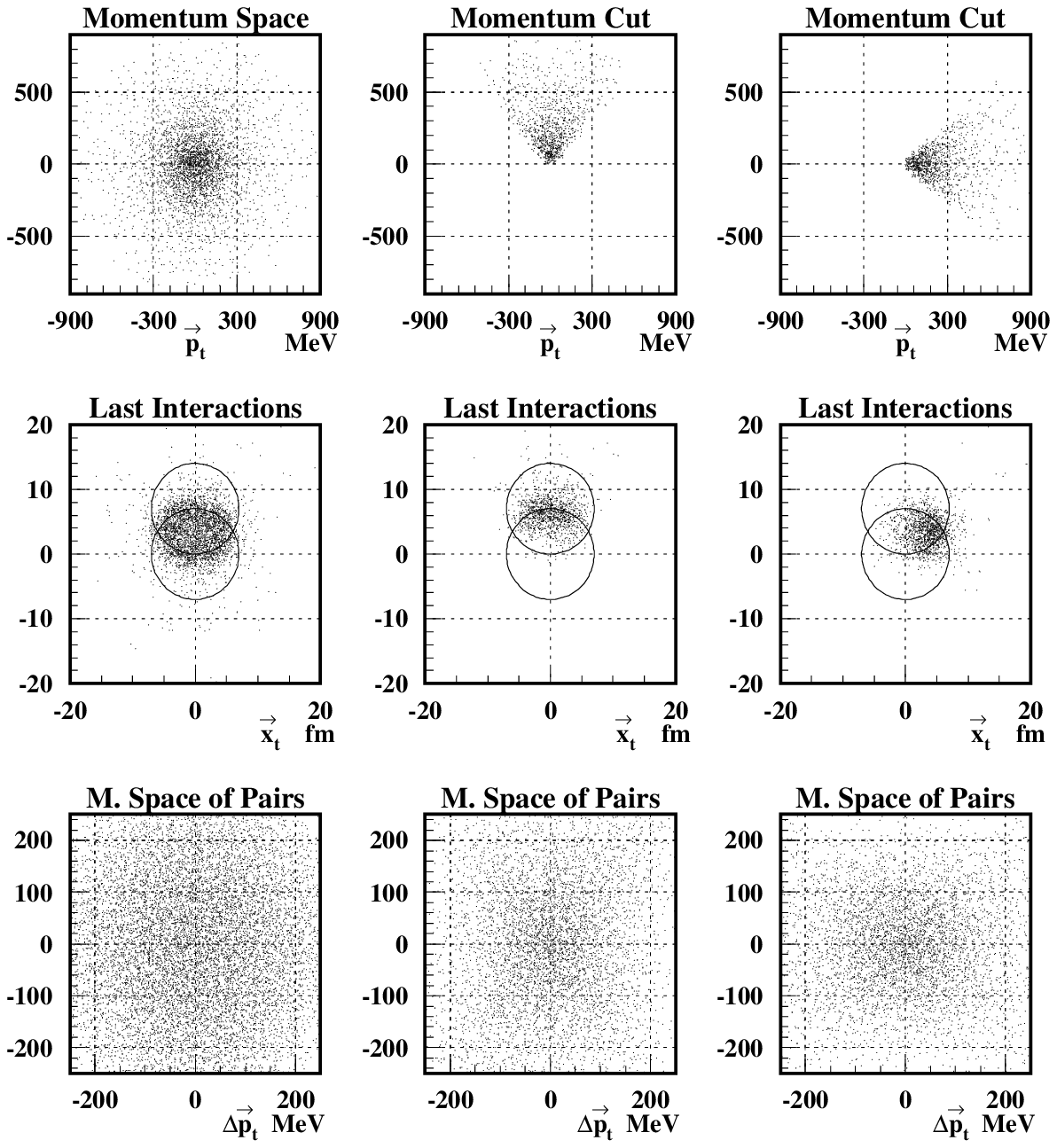}}
\vskip0.05cm
\centerline{\parbox{12cm} {\small {\bf Fig.6 } Directional cuts in
momentum space select pions from different regions of Last Interactions 
distribution. Distribution of pairs in relative momentum space 
($\Delta \vec p_t$) still allows two-dimensional HBT analysis$^3$.
 }}
\vskip0.2cm

Pions with different directions of transversal momentum come from different
regions of $\rho ^{LI} (\vec x, \vec p )$\footnote{Directional cut
in transversal momentum space $(p_x,p_y)$ does not restrict distribution
of pion pairs in relative momentum space $(\Delta p_x, \Delta p_y)$ 
significantly.}. If distribution $\rho ^{LI} (\vec x)$ is asymmetrical
in transversal plane (see Fig.6) then directional cuts in momentum 
space select spatial regions of different size what leads to different
radii parameters extracted by HBT method.
Azimuthally sensitive HBT analysis based on directional momentum cuts was
already performed on AGS data \cite{QM95}.

Common feature of the method based on $\vec x - \vec n_p$ correlation
and X-Y HBT is that orientation of impact parameter is determined
for each event (e.g by transversal flow analysis). 
In the next section we sketch 
a possibility to perform azimuthally sensitive HBT without information about
impact parameter orientation.

\vskip0.25cm
\begin{center}
{\bf 5. Can We See the Reaction Plane via HBT ?}
\end{center}

In experiment the orientation of impact parameter $\vec b$ is random for
non-central collisions. Thus transversal shape of source distribution
will not be visible by HBT if the correlation function
is constructed as a sum of randomly oriented single event contributions.
This situation is similar to transversal flow analysis \cite{Poskanzer}
where event by event
(EbyE) technique allows to determine a presence of azimuthal 
asymmetry \cite{Ollie}.

Here we sketch EbyE HBT technique similar to that 
utilized in transversal
flow analysis \cite{Zhang}. Let us rewrite X-Y decomposition of
transversal correlation function
$$
C^T(\Delta p_x, \Delta p_y)=
1 + e^{-\Delta p^2_x R^2_x/2 -\Delta p^2_y R^2_y/2}
\eqno{(13)}
$$
in spherical coordinate system: 
$$
\Delta p_x=\Delta p_r \cdot \cos(\phi)\quad ; \quad
\Delta p_y=\Delta p_r \cdot \sin(\phi)
\eqno{(14)}
$$
After substitution (14) into (13) we obtain:
$$
C^T(\Delta p_r,\phi ) = 1 + e^{-\frac {\Delta p_r^2}{4}\big[R_0^2 +
\Delta R^2\cos (2\phi)\big]}
\eqno{(15)}
$$
where $R_0^2=R^2_x+R^2_y$ ; $\Delta R^2 = R^2_x - R^2_y$. 
Thus non-zero value
of $\Delta R$ parameter is a signature of transversal asymmetry of 
the correlation
function and consequently also of the source distribution. Random orientation
of impact parameter means a simple random shift in azimuthal angle 
$\phi \longrightarrow \phi + \Delta ^i$ in spherical coordinate system.
Equation (15) can then be rewritten in the form:
$$
C^T(\Delta p_r,\phi )=1+ e^{-\frac {\Delta p_r^2}{4}\big[R_0^2 +
E_x \cos(2\phi) + E_y \sin (2\phi) \big]}
\eqno{(16)}
$$

where parameters $E_x = \Delta R^2\cos (\Delta ^i)$ ; 
$E_y = \Delta R^2\sin (\Delta ^i) $ depend on 
$\Delta ^i$ which is different 
for each event. 
This expression for single-event correlation function is similar to
the second order fourier decomposition of azimuthal distribution $R(\phi )$ 
used in transversal flow analysis \cite{Zhang}:
$$
R(\phi ) = x_o + x_2 \cos (2\phi ) + y_2 \sin (2\phi )
\eqno{(17)}
$$

Non-zero positioned maximum in $v_2=\sqrt {x_2^2 + y_2^2}$ distribution
constructed from many single event fit results is a signature of non-zero
value of average $\bar v_2$ parameter \cite{Zhang}. 
Based on this analogy we imply that
non-zero positioned maximum of $\Delta R^2 = \sqrt {E_x^2 + E_y^2}$ 
distribution
obtained by many single-event HBT fit procedures could be a signature of
asymmetrical shape of $\rho (\vec x)$ distribution in transversal plane.

In this EbyE approach the orientation of impact parameter is not 
determined by other
(e.g. flow) method. Therefore we dare to say that experimentally observed
non-zero average value of $\Delta R$ parameter extracted by this method would
mean observation of the reaction plane via HBT.

We are aware that here we give a very rough description of this method. 
We have not touched other important topics e.g. the construction of 
single-event correlation function or final state interactions correction
\cite{Boal}.

There exists  also a possibility to 
perform HBT search for the reaction plane based on 
$\vec x - \vec n_p$ correlation described in the preceeding section.
This approach would be based on the study of (anti)correlation of the
radii $R_{\phi _1}, R_{\phi _2}$ extracted for e.g. orthogonal directional
cuts in momentum space.

These topics are however beyond the scope of this paper now. In the next 
section we describe shortly a previously reported \cite{Pol,Er} possibility
of further enhancement of HBT analysis.

\begin{center}
\vskip0.3cm
{\bf 6. Inverse Transformation}
\end{center}

General principle of HBT is the following: We measure a result of
interference phenomenon (correlation function $C(\Delta \vec p)$) and
by analysis of $C(\Delta \vec p)$ we reconstruct some information about 
a source emitting bosons. 

This principle is similar to holography methods where however full spatial
information about source can be obtained. Full reconstruction is
allowed by the interference of coherent radiation.
In most of literature about HBT a statement for degenerated (constant)
correlation function  in the case of coherent emission
of bosons can be found. 
However it is not excluded that for coherently emitted bosons 
(produced by e.g. Chiral Condensates \cite{Koch}) a slightly different
-- holographic\footnote{For the inverse transformation based on
Helmholtz-Kirchhoff theorem see \cite{Barton}.}
(\cite{Fadley}) approach to HBT exists.
Nevetherless experimental situation is the main {\it Veto device} which
decides the fate of theoretically existing methods. Therefore
we shall deal only with uncoherently radiated bosons in this section.

As it was shown in work \cite{Pol} for a static (sharp freeze-out)
$\vec x - \vec p$ uncorrelated source we can write:
$$
C(\Delta p_x,\Delta p_y)=\int \int S(D_x,D_y) \big[1+\cos (
\Delta p_x \cdot \Delta D_x + \Delta p_y \cdot \Delta D_y)\big] dD_x dD_y
\eqno{(18)}
$$
where $S(D_x,D_y)$ is a probability distribution for emission of bosons
at a relative distance $\vec D = (D_x,D_y)$. Inverse transformation from 
$C(\Delta p_x,\Delta p_y)$ to $S(D_x,D_y)$ in equation (18) exists.
It is quite tempting to perform this procedure on our simulated CASCUS 
events.

On Fig.7 we show $S(D_x,D_y)$ - symmetrized result of inverse transformation
of correlation function $C(\Delta \vec p_t)$ for artificial S-Pb events
(see Fig. 1a) in X-Y {\bf a)} and Side-Out {\bf b)} decomposition.
Images {\bf c)}, {\bf d)} are result of the same procedure performed
for two-point source (see Fig.2).

\vskip0.2cm
\begin{center}
\begin{tabular}{c c c}
\mbox{\epsfxsize=4.0cm\epsffile{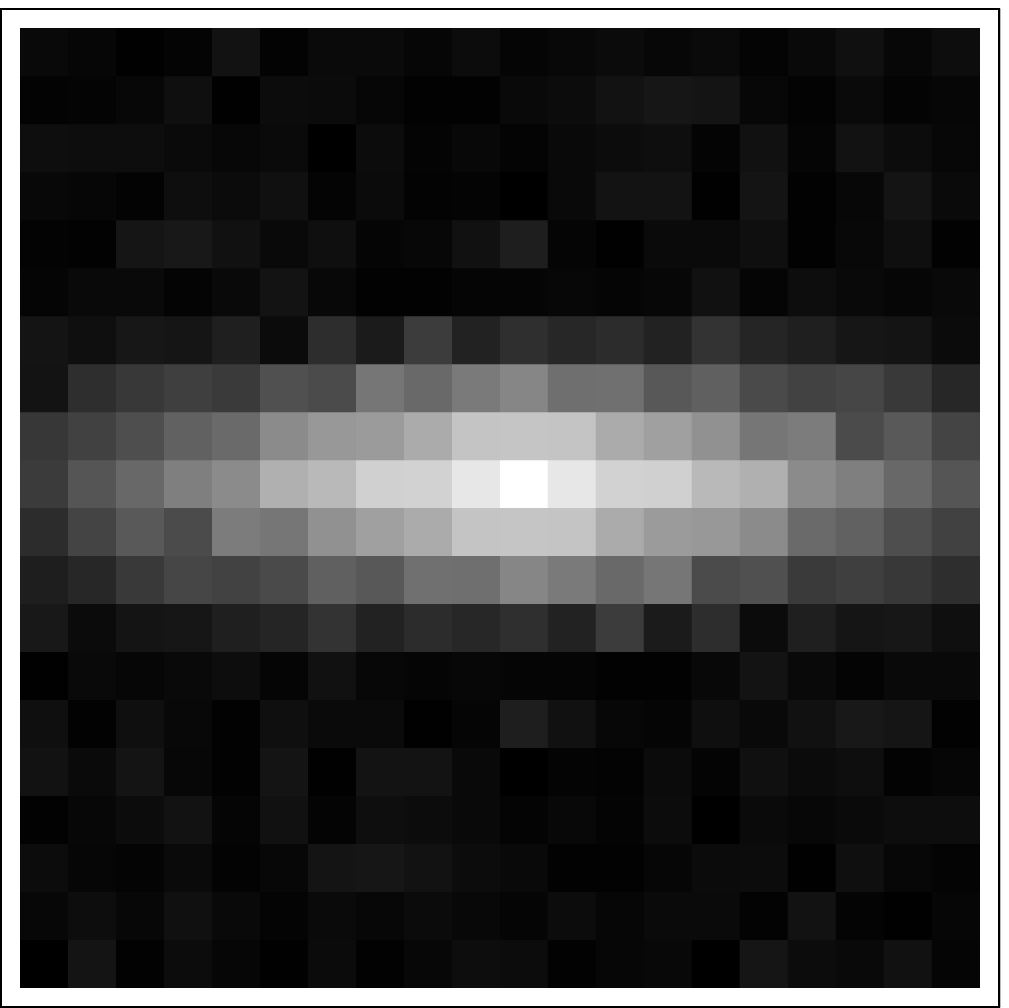}} & &
\mbox{\epsfxsize=4.0cm\epsffile{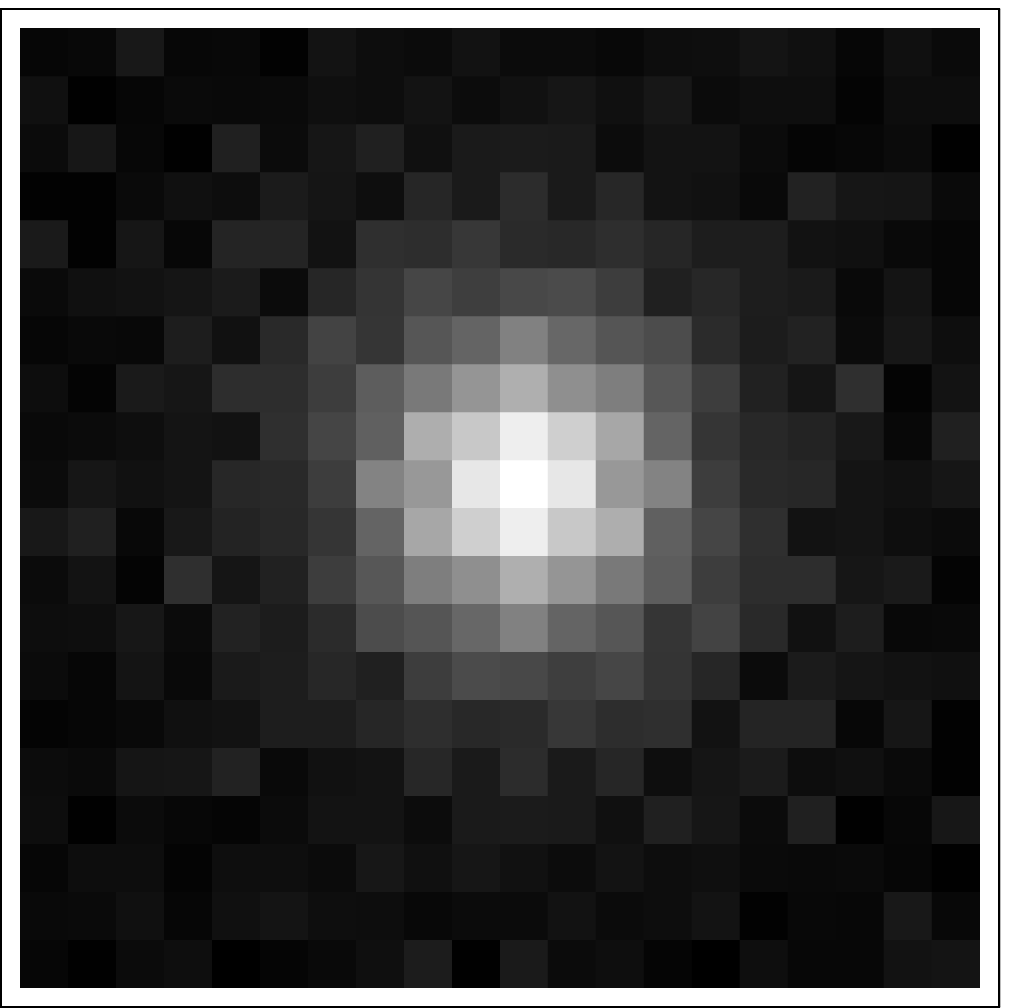}} \\
{\bf a)} &  & {\bf b)} \\
 &  & \\
\mbox{\epsfxsize=4.0cm\epsffile{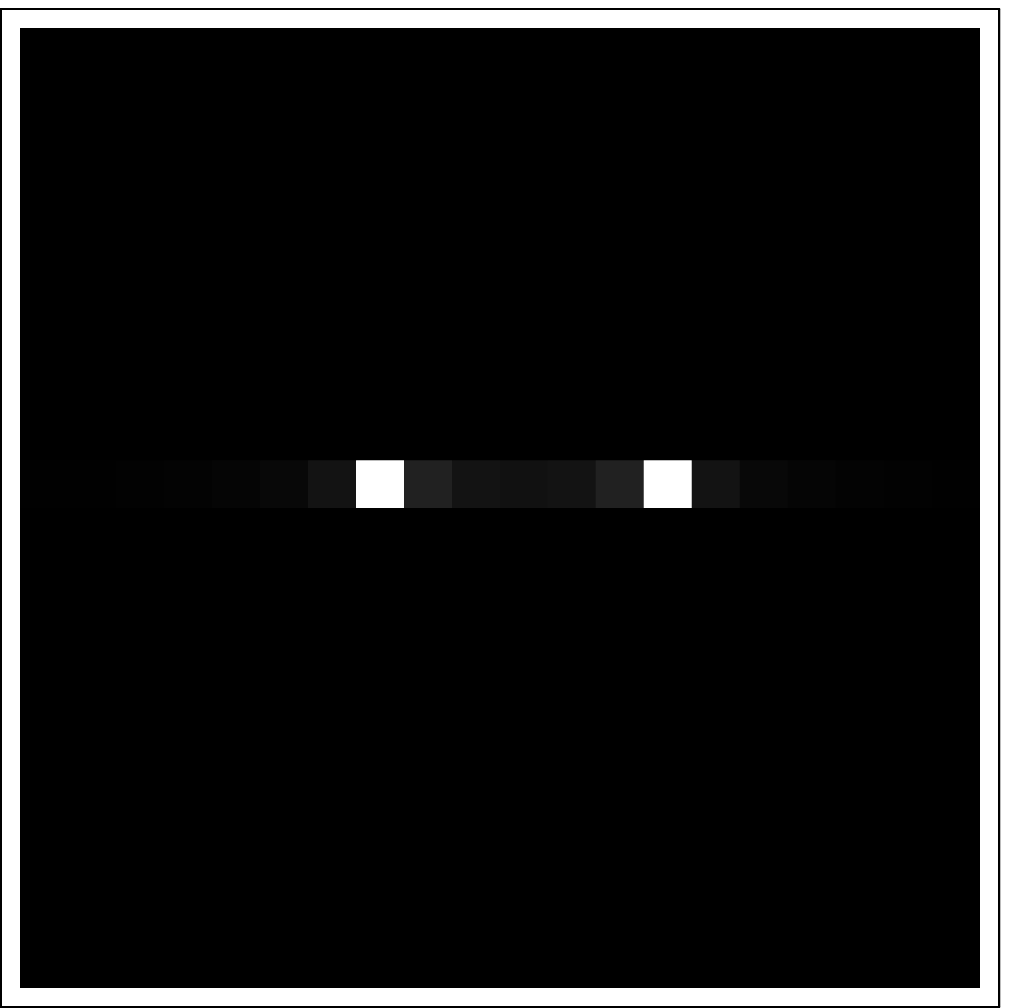}} & &
\mbox{\epsfxsize=4.0cm\epsffile{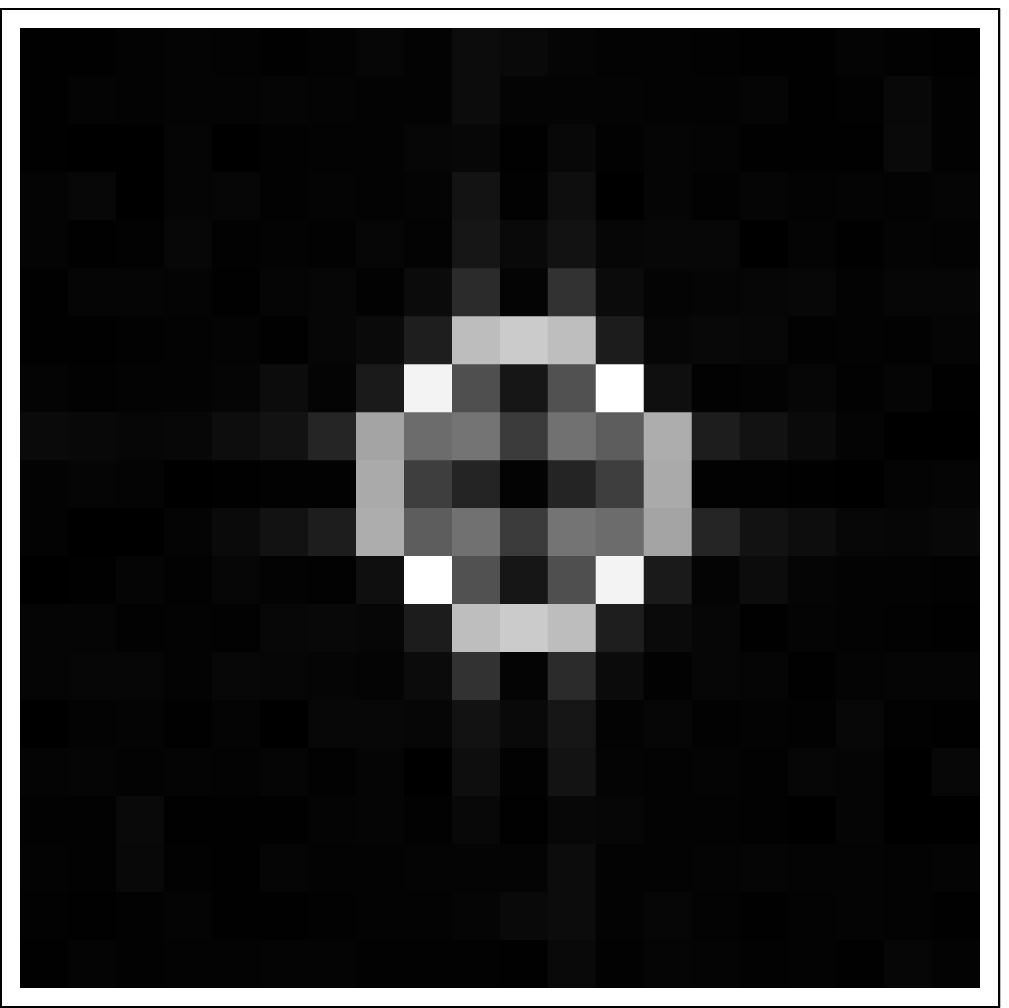}} \\
{\bf c)} & & {\bf d)} 
\end{tabular}
\end{center}
\vskip0.02cm
\centerline{\parbox{12cm} {\small {\bf Fig.7 } Results of inverse 
transformation of equation (18) for 
correlation function of toy S-Pb events in {\bf a)} X-Y (Fig.3b) and {\bf b)}
Side-Out (Fig.1b) decomposition. Images {\bf c)}, {\bf d)} are obtained by
the same procedure for two-point source (Fig.2). Distance between two
dots in X-Y decomposition Image {\bf c)} is 6fm. All images have the same
scale.
 }}
\vskip0.2cm

Distance distribution $S(\vec D)$\footnote{See \cite{Pol,Er} for
the explanation of symmetrized $S(\vec D)$ distribution.}
in X-Y decomposition reflects
clearly shape  of the original source distribution shown
on Fig.3a. This feature allows to study a transversal shape of the source
distribution in non-central collisions. Image Fig.7c) confirms that
X-Y HBT method is sensitive also to a non-trivial structure of $\rho (\vec x)$.
This feature might allow to perform a search for non-statistical inhomogenities
possibly generated by a phase transition in central heavy ion 
collisions \cite{Csernai}.

\newpage
\begin{center}
{\bf Conclusions}
\end{center} 

We have shown that X-Y decomposition HBT analysis is suitable
for the investigation of geometrical properties of the source distribution.
This opens a possibility to study shape  
of the volume radiating bosons
in non-central heavy ion collisions and possibly also non-trivial structure
(inhomogenities) in central heavy ion collisions.

\vskip0.4cm
\begin{center}
{\bf Acknowledgements}
\end{center}

Author is indebted to professor Hans-Georg Ritter and to 
members of NA49 Collaboration for valuable support.

Special thanks are directed to P.Z\'avada for the possibility
to use output of cascading generator \cite{Zavada} and to Anna Nogov\'a for 
help during
the work on rescattering program \cite{Acta}.
This work was supported in part by Lawrence Berkeley National Laboratory,
by QM'96 Heidelberg fund and by VEGA Grant 1001 at Slovak Academy of 
Sciences.

\newpage

\end{document}